\documentclass[conference]{IEEEtran}
\usepackage{amsmath,amssymb,amsfonts}
\usepackage{algorithmic}
\usepackage{graphicx}
\usepackage{textcomp}
\usepackage{xcolor}
\usepackage{balance}
\usepackage{fancyhdr}
\usepackage{url}
\usepackage{caption}
\usepackage{hyperref}
\usepackage[square,comma,sort&compress,numbers]{natbib} 
\usepackage{threeparttable,multirow,adjustbox,threeparttable}
\usepackage{pbalance}  
\usepackage{hyperref}
\usepackage{adjustbox} 
\usepackage{orcidlink}
\usepackage{verbatim}
\hypersetup{
    colorlinks=true,        
    linkcolor=blue,         
    citecolor=blue,         
    filecolor=blue,         
    urlcolor=blue           
}

\usepackage[symbol]{footmisc}

\begin{document}
\title{Pretrained-Guided Conditional Diffusion Models for Microbiome Data Analysis}

\author{
\IEEEauthorblockN{
Xinyuan Shi$^{1,\dag}$, 
Fangfang Zhu$^{2,\dag}$ and 
Wenwen Min$^{1,*}\orcidlink{0000-0002-2558-2911}$ }\\
\IEEEauthorblockA{
$^1$School of Information Science and Engineering, Yunnan University, Kunming 650091, Yunnan, China \\
$^2$College of Nursing Health Sciences, Yunnan Open University, 650599, Kunming, China
}
}
\maketitle
\renewcommand{\thefootnote}{\fnsymbol{footnote}} 
\footnotetext[2]{These authors contributed equally to this work.} 
\footnotetext[1]{Corresponding author: minwenwen@ynu.edu.cn}                          
\thispagestyle{fancy}
\begin{abstract}
Emerging evidence indicates that human cancers are intricately linked to human microbiomes, forming an inseparable connection. However, due to limited sample sizes and significant data loss during collection for various reasons, some machine learning methods have been proposed to address the issue of missing data. These methods have not fully utilized the known clinical information of patients to enhance the accuracy of data imputation. Therefore, we introduce mbVDiT, a novel pre-trained conditional diffusion model for microbiome data imputation and denoising, which uses the unmasked data and patient metadata as conditional guidance for imputating missing values. It is also uses VAE to integrate the the other public microbiome datasets to enhance model performance. The results on the microbiome datasets from three different cancer types demonstrate the performance of our methods in comparison with existing methods. Source code and all public datasets used in this paper are available at Github (\url{https://github.com/wenwenmin/mbVDiT}) and Zenodo (\url{https://zenodo.org/records/13254073}).
\end{abstract}


\section{Introduction}
Cancer is commonly regarded as a complex human genomic disease \cite{gopalakrishnan2018gut,jin2019commensal}. With ongoing research, it has been discovered that the microbiome contributes significantly to certain types of cancer \cite{ma2018gut,gihawi2023major}. Moreover, unique microbial features have been found within and between tissues and blood of most major types of cancer. These studies reveal the potential of microbiomes as tools for cancer diagnosis or treatment. The acquisition of microbiome data primarily relies on two sequencing technologies: the 16S ribosomal RNA (rRNA) amplicon sequencing and the shotgun metagenomic sequencing. The 16S rRNA amplicon sequencing technique quantifies 16S rRNAs, enabling the identification and differentiation of microbes \cite{calle2019statistical}. Shotgun metagenomic sequencing, also referred to as whole-genome sequencing (WGS), involves sequencing all DNA present in a microbiome sample, capturing entire genomes of microbial species as well as host DNA \cite{yu2017metagenomic,li2015microbiome}. 

A key challenge with microbiome data is the presence of numerous zero values (high sparsity), which is a common issue in both major sequencing methods \cite{calgaro2020assessment}, these zero values may arise due to sampling techniques or other factors leading to data gaps. The high sparsity characteristic of microbiome data results in poor performance for down analysis. Additionally, the small sample size and lack of labels in microbiome data are significant factors that hinder the development of deep learning.

Imputation is a widely used technique to recover missing data and facilitate data analysis. In recent years, some methods have been based on GANs (Generative Adversarial Networks) and autoencoders. GANs have become popular in recent years as a method for generating new data through adversarial learning, producing data that conforms to the distribution of real data \cite{deepmicrogen,huang2023scggan}. Additionally, self-supervised deep learning methods and consistency models are applied in related areas \cite{gaziv2022self,song2023consistency}. Similarly, autoencoder-based deep models, with their probabilistic foundations and capability to manage intricate distributions, excel at capturing the underlying representation of the data, rendering them essential for data imputation. These imputation methods have already achieved success in many fields, e.g.,  image and speech reconstruction \cite{rulloni2012large}, imputation of
unmeasured epigenomics datasets \cite{ernst2015large}, and imputation of gene expression recovery in single-cell RNA-sequencing (scRNA-seq) data \cite{deepimpute,autoimpute,dca,scvi}. However, existing methods have not fully utilized clinical information from patients to improve the imputation performance of the model. This results in wastage of data and missed opportunities to uncover potential data insights.

As generative deep models continue to advance, diffusion models are becoming increasingly popular and have been applied in various fields 
\cite{yang2023diffusion,wang2023observed,li2024stmcdi,li2024spadit}. 
Starting from the initial denoising diffusion
probabilistic models (DDPM) \cite{ho2020denoising}, the diffusion models have gradually evolved into score-based diffusion models \cite{song2021score} and are steadily maturing. Meanwhile, the integration of diffusion models with Transformer has provided researchers with more possibilities to fuse diverse information \cite{rombach2022high}. 

To address the challenges in this field, we introduce a novel method, mbVDiT for microbiome data imputation. Our approach primarily combines the excellent capabilities of variational autoencoders (VAE) in learning latent distributions and diffusion models in generating new data that conforms to the real data distribution. To address the issue of unlabeled microbiome data, we drew inspiration from recent research in the field \cite{he2022masked,hou2022graphmae}. We masked a portion of the original data, which remained invisible throughout the entire modeling process and served as labels for evaluating the imputation performance after filling. After the masked data is encoded into the latent space by the VAE encoder, we re-mask the latent space and employ a conditional score-based diffusion model for training the denoising network. Our goal is to reconstruct a cancer microbiome data matrix by integrating various types of metadata information from patients and handling the label-less nature of the data. Make the most of existing conditions to improve the accuracy of imputed data while preserving the inherent characteristics and distribution patterns of the data.
The main contributions of our proposed method are:
\begin{itemize}
    \item Through the application of pre-training strategy, we achieved common features across different cancer type datasets. Using data from other types of cancers has improved the model performance through cross-utilization caused by the limited number of samples in individual datasets.
    \item Using masking for self-supervised learning addresses the challenge of lacking labels in microbiome data.
    \item Using various types of metadata from cancer patients as conditions has improved the model performance.
\end{itemize}
\begin{figure*}[htp]%
\centering
\includegraphics[width=1\textwidth]{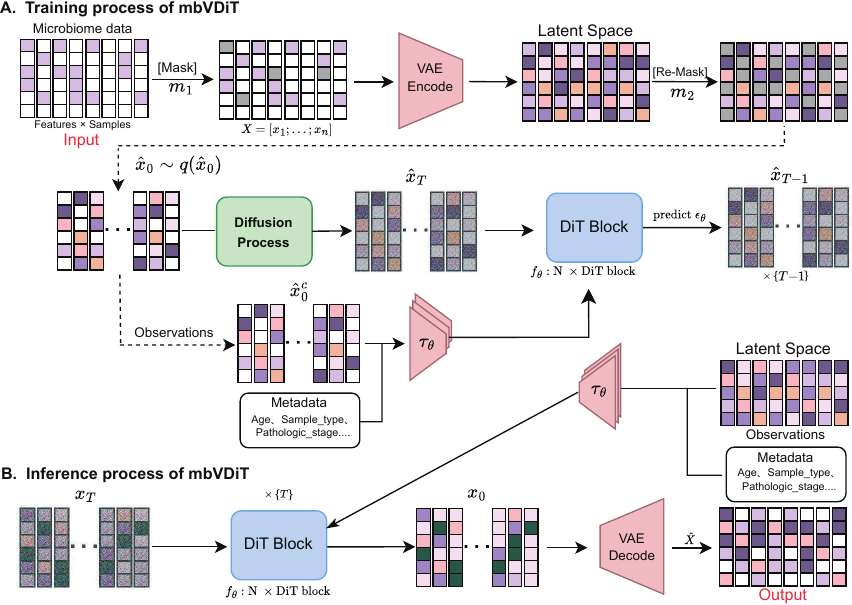}
\caption{The architecture of the proposed mbVDiT, where a VAE-based pre-trained model is used to integrate other public microbiome datasets to enhance model performance. 
(A) Training process: The latent space data $\hat{x}_{0}$ after masking is subjected to noise to obtain $\hat{x}_{T}$. Then, a denoising network $f_{\theta}$ is trained to predict the noise under the condition. (B) Inference process: The trained denoising network $f_{\theta}$ denoises the random noise $x_{T}$ step-by-step under the guidance of the condition. The data after noise removal is then passed through a decoder, resulting in the predicted imputation for the microbiome data. }
\label{fig1}
\end{figure*}
\section{PROPOSED METHODS}\label{MATERIALS AND MRTHODS}
\subsection{Overview of the proposed mbVDiT}
We propose a conditional diffusion model, mbVDiT, for microbiome data imputation (\autoref{fig1}). The model is able to perform accurate and stable imputation using the unmasked latent microbiome data and patient metadata. 

\subsection{mbVDiT model}
\subsubsection{Mask strategy}\label{Dataset}
Due to the lack of labels in microbiome data, we employ self-supervised learning methods \cite{he2022masked}. To assess the impact of different missing rates on model performance, we randomly mask 10\%, 30\% and 50\% of the non-zero elements for each given data matrix and designate this masked portion as the test set for subsequent model computation and evaluation of performance metrics. Specifically, a mask matrix $M$ is defined.
\begin{equation}
M=\left\{\begin{matrix} 1 ,& \text{ if } x_{ij} \text{ is masked }\\ 0, & \text{ otherwise }\end{matrix}\right.
\end{equation}
The mask matrix has the same dimensions as the input data, where $M\in R^{n\times d} $, $x_{ij}$ represents the masked data. After masking, the final matrix $Y\odot M$ serves as the input for mbVDiT.


\subsubsection{Pre-training VAE}\label{}
The VAE model consists of an encoder and a decoder, maintaining consistency in both structure and functionality with the traditional model. Our main purpose in employing the VAE model is to capture the latent distribution and representation of data features from other cancer type datasets. 

Specifically, we pre-train using four different types of cancer datasets and transfer the resulting pre-trained parameters to the VAE module of mbVDiT. After that, we train mbVDiT using a single dataset. It is important to note that this individual dataset is not included in the four pre-training datasets.

Formally, given input data $\mathit{X}$, $X\in R^{n\times d}$, representing a set of ${n}$ samples with ${d}$ features. the encoder maps the input to a potential distribution, which contains two parameters, variance $\mu _{i }$ and mean $\sigma _{i}^{2}$:
\begin{equation}
    [\mu _{i },~\sigma _{i}^{2}]= Encoder(X_{i} )
\end{equation}
where $\mathit{i}$ denotes the microbial data features of a sample.

Then the decoder obtains a sample $\mathit{Z}$, $Z\sim \mathcal{N} (\mu _{i},\sigma _{i}^{2}  )$, from this latent distribution and completes the reconstruction of the data based on this sample:

\begin{equation}
    \widehat{X} _{i} =Decoder(Z)
\end{equation}

\subsubsection{VAE-encoder in mbVDiT}\label{VAE}
When the number of features exceeds the number of samples, a potential issue known as the curse of dimensionality may arise, which could lead to overfitting during model training and result in decreased model performance. Therefore, we introduce a VAE encoder to reduce the dimensionality of the data and train the model in the latent space containing the underlying distribution of the data. It is worth noting that the VAE module in this model is pretrained, including encoder and decoder.

Specifically, given a masked data $X$, it is mapped to a latent space $H$ containing the underlying distribution of the data through the VAE encoder. By employing a fully connected layer, the encoder is denoted as $f_{\theta}^{en}$. This can be formulated as:
\begin{equation}
    H=f_{\theta }^{en}(X)
\end{equation}
where $H\in R^{n\times l} $, $n$ represents the number of samples and $l$ represents the dimensionality of the latent space. This latent representation will be used for training the diffusion model.

\subsubsection{Conditional score-based diffusion model for posterior estimation}\label{Diffusion}
In order to gain a deeper understanding of the diffusion model, we carefully studied the classic literature in the field of diffusion models \cite{ho2020denoising,sohl2015deep}. Follow the notations in denoising diffusion probabilistic models (DDPM). DDPMs are latent variable models consisting of two main processes: the forward process and the reverse process. The forward process is characterized by a Markov chain, outlined as follows:
\begin{equation}
\begin{aligned}
q(\text{x}_{1:T} \mid \text{x}  _{0} )=&\prod_{t=1}^{T} q(\text{x}_{t} \mid \text{x}  _{t-1} ) \\
q(\text{x}_{t} \mid \text{x}  _{t-1} )=\mathcal{N} (\text{x}_{t} &;\sqrt{1-\beta _{t} }\text{x} _{t-1} ,\beta _{t}\textbf{I} ) \\
 \end{aligned}
\end{equation}
where $\beta _{t}$ is a small positive constant indicative of a noise level, $\left \{ \beta _{t}  \right \} _{t=1}^{T} \in (0,1)$, and the numerical value increases with time. Through reparameterized sampling, we can obtain $q(\text{x}_{t} \mid \text{x} _{0})=\mathcal{N} (\text{x}_{t};\sqrt{\bar{\alpha } _{t}} \text{x} _{0},(1-\bar{\alpha } _{t} )\textbf{I} )$, where $\hat{\alpha } _{t} =1-\beta _{t} $ and $\alpha _{t}$ is  the cumulative product $\bar{\alpha } _{t} = {\textstyle \prod_{i=1}^{t}} \hat{\alpha } _{i} $. As a result, $x_{t}$ can be expressed by the equation $\text{x} _{t} =\sqrt{\bar{\alpha } _{t} } \text{x}_{0} +(1-\bar{\alpha } _{t} )\epsilon $, with $\epsilon \sim \mathcal{N} (0,\textbf{I} )$. On the contrary, the reverse process is a denoising process, obtaining $x_{0}$ from $x_{t}$. This process can be defined as a Markov chain:
\begin{equation}
\begin{aligned}
 p_{\theta } (\text{x} _{0:T} )=p(\text{x}_{T}&  )\prod_{t=1}^{T} p_{\theta } (\text{x} _{t-1} \mid \text{x} _{t} ) \\
 \text{x}_{T} \sim& \mathcal{N} (0,\textbf{I} )
\end{aligned}
\label{equ:4}
\end{equation}

\begin{equation}
\begin{aligned}
p_{\theta } (\text{x}_{t-1}  \mid \text{x} _{t} )=&\mathcal{N} (\text{x}_{t-1} ;\mu _{\theta } (\text{x}_{t},t )  ,\sigma ^{2}_{\theta  } (\text{x} _{t} ,t)\textbf{I} ) \\
 \mu _{\theta } (\text{x}_{t},t )=&\frac{1}{\bar{\alpha } _{t} } (\text{x}_{t}  -\frac{\beta_{t}}{\sqrt{1-\bar{\alpha }_{t} } }\epsilon _{\theta } (\text{x} _{t},t ) ) \\
 \sigma_{\theta  } &(\text{x} _{t} ,t)=\beta _{t}^{1/2}
 \end{aligned}
 \label{equ:5}
\end{equation}
where $\beta _{t} =\left\{\begin{matrix}  \frac{1-\bar{\alpha } _{t-1} }{1-\bar{\alpha } _{t} } \beta _{1} ,& \text{for } t>1 \\  \beta _{1}, &\text{for } t=1\end{matrix}\right.$ and $\epsilon _{\theta } (\text{x}_{t} ,t)$ is a trainable denoising function. In our proposed approach, we utilize a score-based diffusion model, which distinguishes itself from the conventional DDPM primarily through distinct coefficients employed during the sampling phase.

\subsubsection{Conditioning mechanisms}\label{Condition}
Our proposed method intends to estimate the entire dataset by leveraging the known data and patient metadata, such as pathological etc, as prior conditions. We aim to utilize the known data to complete missing value imputation, thereby reconstructing a complete representation of the dataset. 

\textbf{(1) Observable data as the condition}
We denote the observed data as $x_{0}^{c} $, $x_{0}^{c} =(E_{n,l} -m_{2})\odot x_{0} $, where $x_{0}$ represents the latent representation at time step 0, $m_{2}$ is an element-wise indicator, $m_{2}\in \left \{ 0,1 \right \} ^{n\times l} $, where ones denote masked values and zeros denote observed values, and $E_{n,l} $ is all-ones matrix \cite{li2024stmcdi}. 

\textbf{(2) Patient metadatas as the condition}
Patient metadata includes both textual and floating-point data. For the integration of multiple types of conditions, we referred to some literature \cite{takagi2023high,kim2022diffusionclip,ho2022classifier}. To pre-process $y_{i}$ originating from diverse modalities, we introduce multiple domain-specific encoders (MLP) $\tau _{\theta ,i}$ that map $y_{i}$ to an intermediate representation, the dimensionality of this representation is the same as that of the feature dimension.

Overall, the conditions of the diffusion model are composed of observable data and metadata, which is then mapped to the intermediate layers of the backbone via a cross-attention layer implementing.
\begin{equation}
C=\tau(x_{0}^{c}) +\tau _{\theta ,i} (y_{i} )
\end{equation}
where $i$ represents the number of types of patient metadata and $\tau(\cdot )$ is a multilayer perceptron
(MLP) that converts unmasked part of data as an input condition.

We define the imputation data for each time step as $x_{t}^{*} $. Hence, the conditional mechanisms within mbVDiT aim to estimate the probabilistic:
\begin{equation}
p_{\theta } (x_{t-1}^{*} |x_{t}^{*} ,C )
\end{equation}

To enhance the utilization of observed values as prior condition for the diffusion model in performing missing value imputation, we reformulate Equation (\ref{equ:4}) and Equation (\ref{equ:5}) as:
\begin{equation}
\begin{aligned}
 p_{\theta } (x_{0:T}^{*} |C )=p(x_{T}^{*}& )\prod_{t=1}^{T} p_{\theta } (x_{t-1}^{*} |x_{t}^{*},C) \\
 x_{T}^{*} \sim& \mathcal{N} (0,\textbf{I} )
\end{aligned}
\end{equation}
\begin{equation}
p_{\theta } (x_{t-1}^{*} |x_{t}^{*} ,C )=\mathcal{N} (x_{t-1}^{*} ;\mu _{\theta }(x_{t}^{*} ,t|C ), \sigma _{\theta }^{2}(x_{t}^{*} ,t|C )\textbf{I})
\end{equation}

The $\tau _{\theta } $ part is a multi-layer perceptron (MLP), aiming to map the observed data to the input condition $\tau _{\theta } (x_{0}^{c} )$. Subsequently, the condition is mapped to the intermediate layers of the backbone network through a cross-attention layer to guide the model during training. Attention layer can be formalized as:
\begin{equation}
\begin{aligned}
 Attention(Q,K,V)=softmax(\frac{QK^{T} }{\sqrt{d} } )\cdot V \\
 Q=W_{Q}^{} \cdot \varphi (x_{T}^{*} ),K=W_{K} \cdot \tau _{\theta } (C ),V=W_{V} \cdot \tau _{\theta }(C )
\end{aligned}
\end{equation}
where $\varphi (x_{T}^{*} )$ denotes a intermediate representation of the backbone network implementing $\epsilon _{\theta } $ and $W_{Q}$, $W_{K}$, $W_{V}$ are learnable projection metrices.

\subsubsection{Inference process of mbVDiT}\label{}
During the inference phase, we randomly sample complete noise data directly from a standard normal distribution and apply the denoising process. Finally, we decode the new samples from the denoised latent distribution.

\subsubsection{Training process and Loss function}\label{Loss}
Training mbVDiT involves two stages: (1) pre-training the VAE encoder and decoder; and (2) training the diffusion model in the latent space.

The loss function in the first stage consists of two parts, reconstruction loss and KL divergence respectively. Simultaneously training the encoder and decoder to minimize this loss.
\begin{equation}
    \mathcal{L}_{VAE} = \parallel X -\widehat{X}  \parallel  +KL(\mathcal{N}\sim (\mu  ,\sigma ^{2} ) \parallel \mathcal{N} (0,1))
\end{equation}

For the second stage, under the condition of $C $ and imputation targets $x_{t}^{*} $, we proceed to refine $\epsilon _{\theta } $ by minimizing the ensuing loss function:
\begin{equation}
L_{DM} = \mathbb{E} _{x_{0} \sim q(x_{0} ),\epsilon \sim \mathcal{N}(0,\textbf{I} ) ,t} \left \| \epsilon -\epsilon _{\theta } (x_{t}^{*},t\mid C ) \odot M\right \|_{2}^{2} 
\end{equation}

\subsection{Evaluation metrics}\label{Evaluation metrics}
To evaluate the performance of mbVDiT and baseline methods, we used four evaluation metrics on three datasets: Pearson correlation coefficient (PCC), Cosine distance (Cosine), Root Mean Square Error (RMSE), and Mean Absolute Error (MAE).
\begin{table}[htp]
\caption{The list of microbial datasets used in the study includes the three cancer types with the highest number of samples from TCGA.}\label{tab1}
\begin{adjustbox}{width=.492\textwidth}
\fontsize{12}{12}\selectfont    
\begin{tabular}{l|cc|cc|c}
\hline
\multicolumn{1}{c|}{\multirow{2}{*}{Datasets}} & \multicolumn{2}{c|}{Number of Samples/Microbes}        & \multicolumn{2}{c|}{Prepro. Samples/Microbes}         & \multirow{2}{*}{Droupout Rate} \\ \cline{2-5}
\multicolumn{1}{c|}{}                          & \multicolumn{1}{c|}{\#Samples} & \#Microbes            & \multicolumn{1}{c|}{\#Samples} & \#Microbes           &                                \\ \hline
STAD                                           & \multicolumn{1}{c|}{530}       & \multirow{3}{*}{1289} & \multicolumn{1}{c|}{530}       & \multirow{3}{*}{106} & 87.61\%                        \\
COAD                                           & \multicolumn{1}{c|}{561}       &                       & \multicolumn{1}{c|}{561}       &                      & 63.20\%                        \\
HNSC                                           & \multicolumn{1}{c|}{587}       &                       & \multicolumn{1}{c|}{587}       &                      & 79.63\%                        \\ \hline
READ                                           & \multicolumn{1}{c|}{182}       & \multirow{2}{*}{1289} & \multicolumn{1}{c|}{182}       & \multirow{2}{*}{106} & 67.06\%                        \\
ESCA                                           & \multicolumn{1}{c|}{248}       &                       & \multicolumn{1}{c|}{248}       &                      & 88.99\%                        \\ \hline
\end{tabular}
\end{adjustbox}
\end{table}
\begin{table*}[htp]
\centering
\caption{Performance comparison between mbVDiT and baselines on three microbiome datasets (STAD, COAD and HNSC) using four evaluation metrics. The black bold part represents the best performance.}
\label{tab2}
\begin{adjustbox}{width=1\textwidth}
\fontsize{9}{8}\selectfont          
\setlength{\arrayrulewidth}{0.05mm}  
\begin{tabular}{l|c|c|c|c|c|c|c|c}
\hline
\textbf{PCC $\uparrow$}  & KNN         & DeepImpute \cite{deepimpute} & AutoImpute \cite{autoimpute}  & DCA \cite{dca}         & CpG \cite{cpg}         & scVI \cite{scvi}       & DeepMicroGen \cite{deepmicrogen}& \textbf{mbVDiT}                             \\ \hline
\textbf{STAD} & 0.232±0.021 & 0.570±0.067 & 0.477±0.048 & 0.505±0.025 & 0.520±0.049 & 0.596±0.065 & 0.588±0.065  &  \textbf{0.634±0.032} \\
\textbf{COAD} & 0.188±0.033 & 0.653±0.054 & 0.639±0.085 & 0.632±0.062 & 0.599±0.069 & 0.667±0.059 & 0.675±0.049  &  \textbf{0.704±0.062} \\
\textbf{HNSC} & 0.247±0.038 & 0.592±0.032 & 0.550±0.043 & 0.594±0.056 & 0.585±0.017 & 0.559±0.028 & 0.604±0.061  & \textbf{0.626±0.060} \\ \hline
\hline
\textbf{Cosine $\uparrow$} & KNN         & DeepImpute \cite{deepimpute} & AutoImpute \cite{autoimpute}  & DCA \cite{dca}         & CpG \cite{cpg}         & scVI \cite{scvi}       & DeepMicroGen \cite{deepmicrogen}& \textbf{mbVDiT}                             \\ \hline
\textbf{STAD}   & 0.430±0.014 & 0.773±0.059 & 0.787±0.027 & 0.746±0.036 & 0.746±0.027 & 0.775±0.072 & 0.775±0.074  &\textbf{0.806±0.057} \\
\textbf{COAD}   & 0.241±0.037 & 0.769±0.057 & 0.650±0.124 & 0.724±0.064 & 0.688±0.062 & 0.716±0.052 & 0.772±0.072  &\textbf{0.791±0.080} \\
\textbf{HNSC}   & 0.357±0.031 & 0.779±0.044 & 0.794±0.050 & 0.789±0.026 & 0.778±0.015 & 0.776±0.024 & 0.786±0.031  &\textbf{0.802±0.062} \\ \hline
\hline
\textbf{RMSE $\downarrow$} & KNN         & DeepImpute \cite{deepimpute} & AutoImpute \cite{autoimpute}  & DCA \cite{dca}         & CpG \cite{cpg}         & scVI \cite{scvi}       & DeepMicroGen \cite{deepmicrogen}& \textbf{mbVDiT}                             \\ \hline
\textbf{STAD} & 3.371±0.124 & 1.572±0.082 & 1.521±0.055 & 1.629±0.044 & 1.462±0.075 & 2.478±0.085 & 1.469±0.049  & \textbf{1.320±0.053} \\
\textbf{COAD} & 4.336±0.106 & 1.211±0.076 & 1.370±0.070 & 1.183±0.059 & 1.127±0.045 & 2.351±0.057 & 1.002±0.100  & \textbf{0.934±0.054} \\
\textbf{HNSC} & 4.128±0.083 & 1.258±0.058 & 1.364±0.049 & 1.246±0.034 & 1.169±0.022 & 2.168±0.077 & 1.245±0.047  & \textbf{1.155±0.027} \\ \hline
\hline
\textbf{MAE $\downarrow$}  & KNN         & DeepImpute \cite{deepimpute} & AutoImpute \cite{autoimpute}  & DCA \cite{dca}         & CpG \cite{cpg}         & scVI \cite{scvi}       & DeepMicroGen \cite{deepmicrogen}& \textbf{mbVDiT}                             \\ \hline
\textbf{STAD} & 3.255±0.095 & 1.388±0.061 & 1.186±0.039 & 1.234±0.047 & 1.308±0.071 & 2.221±0.068 & 1.325±0.058  & \textbf{0.956±0.050} \\
\textbf{COAD} & 3.862±0.114 & 0.804±0.065 & 0.864±0.128 & 0.728±0.061 & 0.739±0.063 & 2.132±0.049 & 0.627±0.109  & \textbf{0.530±0.070} \\
\textbf{HNSC} & 3.679±0.096 & 0.986±0.064 & 0.981±0.045 & 0.993±0.018 & 0.906±0.021 & 0.847±0.083 & 0.978±0.028  & \textbf{0.807±0.020} \\ \hline
\end{tabular}

\end{adjustbox}
\end{table*}




\section{EXPERIMENTAL RESULTS}\label{result}
\subsection{Dataset and pre-processing}\label{Dataset and pre-processing}
All cancer microbiome datasets with different cancer types were obtained from the TCGA public database \cite{weinstein2013cancer, dohlman2021cancer}. 
Due to the limited data availability, we selected these microbiome datasets from three different cancer types with relatively larger sample sizes as our experimental datasets including
Stomach adenocarcinoma (STAD),
Colon adenocarcinoma (COAD) and Head and Neck Squamous Cell Carcinoma (HNSC), specifically, Esophageal carcinoma (ESCA) and Rectum adenocarcinoma (READ) datasets with smaller sample sizes were used for each pre-training, the detailed description of the relevant data is provided in \autoref{tab1}.

For data preprocessing, we first conducted screening on microbial features. To maintain consistency, we then combined the three datasets. Then, microbes with missing data rates (dropout) exceeding 95\% were removed. mbVDiT requires data from each sample to be on the same scale, so the next step is to normalize the data. 

However, properly normalizing microbial data poses a challenge. We have selected a normalization method from benchmark papers \cite{cappellato2022investigating,jiang2021mbimpute}. For each distinct cancer dataset, Given a matrix $M$, mbVDiT normalizes samples row-wise. The normalized matrix is denoted as $M_{ij}^{'} \in R^{n\times d}$, where $M_{ij}^{'} =10^{2} \frac{M_{ij} }{ {\textstyle \sum_{k=1}^{d}} M_{ik} } $. To mitigate the influence of extremely large counts, we applied the logarithmic transformation to the normalized matrix. The log-transformed matrix is denoted as $Y \in R^{n\times d}$, where $Y =\log_{10}{(M_{ij}^{'} +  1.0)} $.

\subsection{Baseline methods}\label{Baseline methods}
To evaluate the performance of mbVDiT, we first compared with KNN imputation. Due to the limited methods for imputing microbiome data, we sought high-quality alternatives. Although we reproduced this particular method \cite{jiang2021mbimpute}, it ran too long and produced suboptimal results, so we excluded it from the baseline comparison. Additionally, we compared several state-of-the-art imputation methods for single-cell data:
\begin{itemize}
    \item DeepMicroGen \cite{deepmicrogen}: It is a generative adversarial network-based method for longitudinal microbiome data imputation.
    \item DeepImpute \cite{deepimpute}: It is an accurate, fast and scalable deep neural network method to impute single-cell RNA-seq data.
    \item AutoImpute \cite{autoimpute}: It is a autoencoder based method imputation of single-cell RNA-seq data.
    \item DCA \cite{dca}: It is a method for denoising single-cell RNA-seq data using a deep count autoencoder.
    \item CpG \cite{cpg}: It is a transformer-based method for imputation of single-cell methylomes.
    \item scVI \cite{scvi}: It is deep generative modeling for single-cell transcriptomics.
\end{itemize}

The detailed results are presented in \autoref{tab2}.

\begin{figure}[htp]%
\centering
\includegraphics[width=0.5\textwidth]{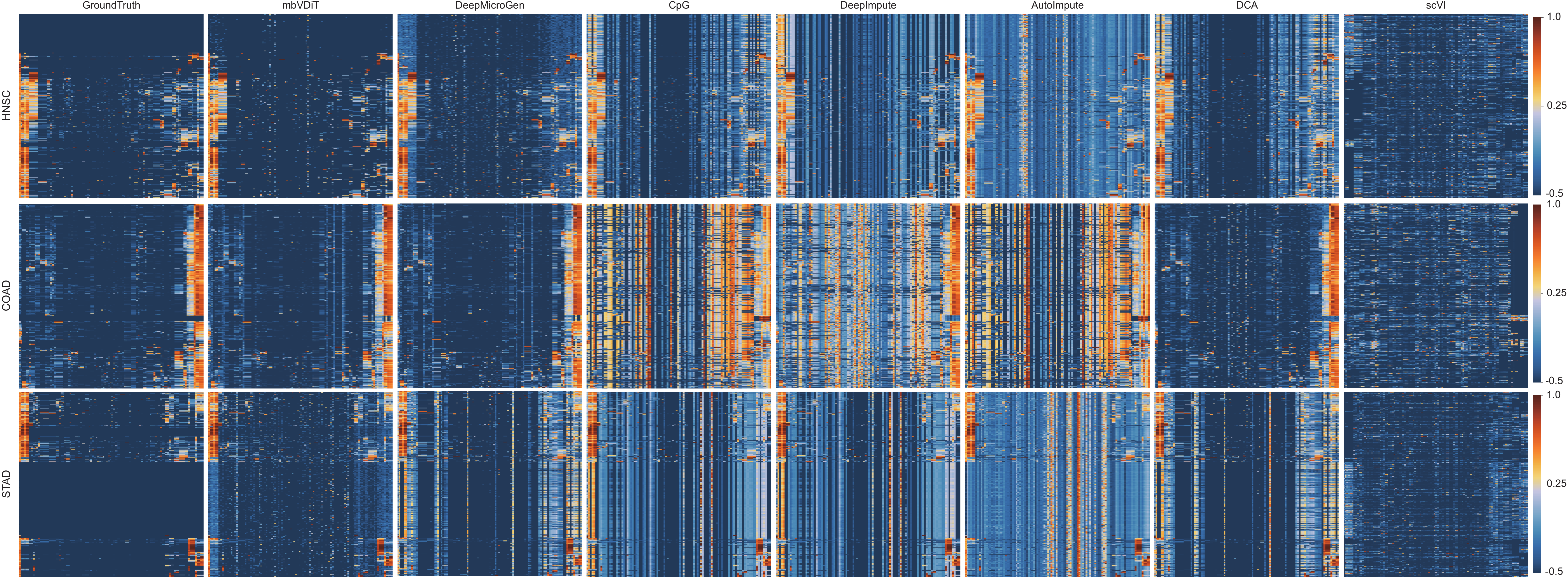}
\caption{Visualization of the imputation performance of various baseline methods.}
\label{fig2}
\end{figure}

\begin{figure}[htp]%
\centering
\includegraphics[width=1\columnwidth]{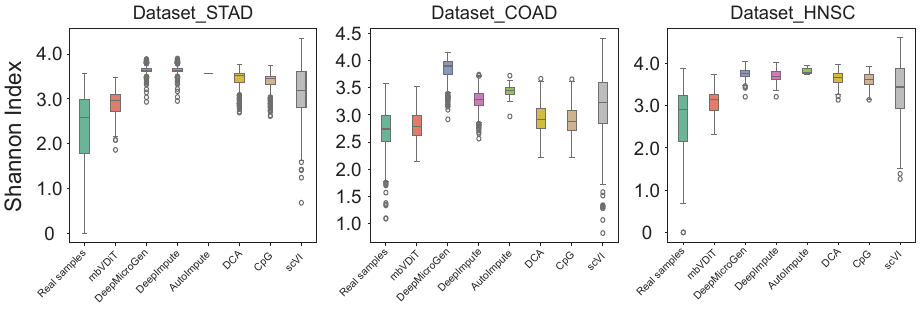}
\caption{Comparison of alpha-diversity based on Shannon index measured from real data and the imputed data by different methods.}
\label{fig3}
\end{figure}

\subsection{mbVDiT improves the accuracy of imputed data}
To further demonstrate the superior performance of our method compared to other state-of-the-art methods, we visualize the imputed microbiome data for each method (\autoref{fig2}). The closer the predicted heatmap is to the ground truth, the higher the prediction accuracy. We splice the mask part to form the real label, and we apply the same method to the imputation results of each baseline method. 

We normalized both the actual expression matrix and the imputed matrix. To enhance the visualization, we used hierarchical clustering to sort the microbes, ensuring that the expression matrix exhibits pattern-like features. 

From the figure (\autoref{fig2}), it is evident that the hierarchical clustering diagram obtained using our method for data imputation shows the most distinct hierarchy and the data patterns are closest to the real data patterns. This further demonstrates that our method provides the best imputation results and effectively preserves the original data distribution.

\subsection{mbVDiT preserves the similar characteristics of missing samples in data imputation}
To further validate the performance of methods, we conducted an alpha-diversity analysis on the imputed microbiome data. Firstly, we calculate the relative abundance of each species for each sample, and then we calculated the alpha-diversity using the Shannon index for both the real samples and the imputed data, and a box plot of alpha diversity was also plotted (\autoref{fig3}). We masked 30\% of the original data and performed imputation on this basis, then calculated the alpha-diversity values of the imputed data. 

We observed that the alpha diversity box plot calculated from the imputed data using our method is the closest to that of the original data, while other methods show relatively larger deviations. Our proposed method better preserves the community distribution characteristics and species abundance of the original data after imputing missing values. The microbial diversity after imputation closely matches the real data for each cancer type.



\subsection{mbVDiT retains the Pearson correlation coefficients of imputed data}
\begin{figure}[htp]%
\centering
\includegraphics[width=0.5\textwidth]{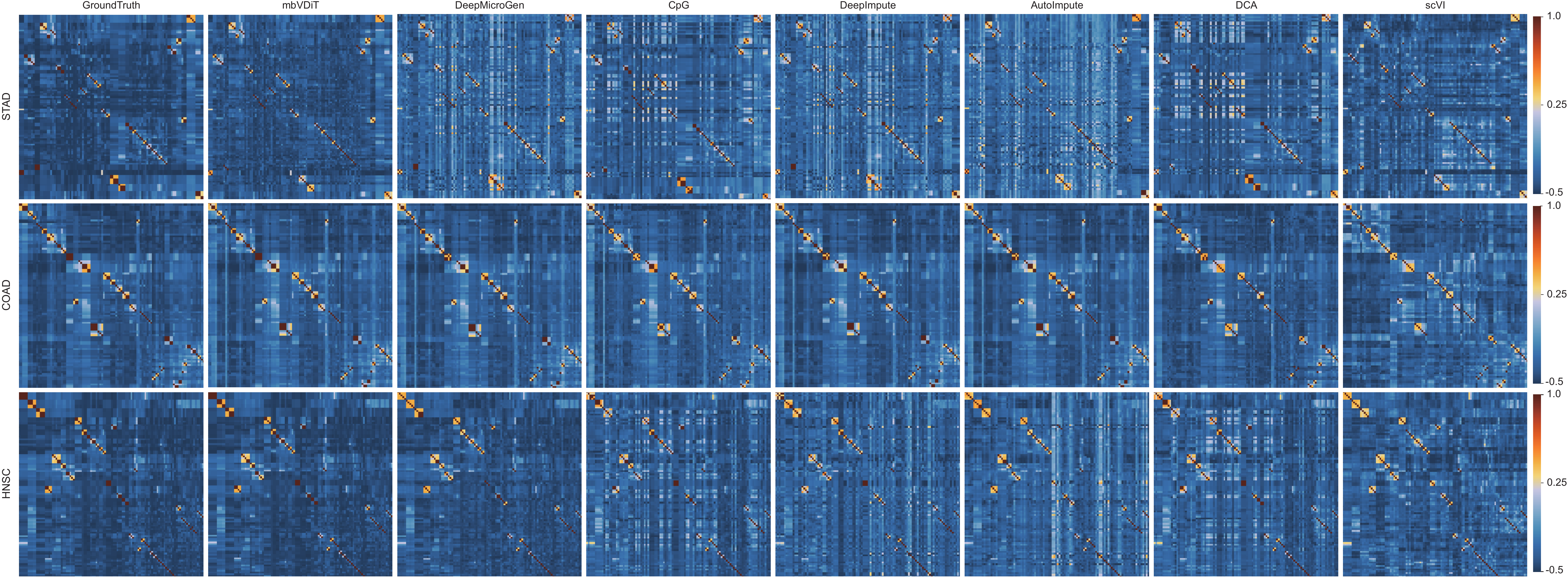}
\caption{Visualization of Pearson correlation coefficients for imputed data between microbe and microbe.}
\label{fig4}
\end{figure}

To further validate the superior performance of our model. Here, we calculated the Pearson correlation coefficients for imputed data. First, we calculated the PCCs between each pair of microbes in the imputed microbiome matrices for each method. After obtaining the correlation coefficient matrices, we applied hierarchical clustering to sort the matrices, enhancing the visualization effect (\autoref{fig4}).

It can be observed that our method preserves microbial distribution characteristics similar to real samples after imputation. Among the baseline methods, DeepMicroGen has the best visualization effect, and its visualization effect of PCCs between microbes is very similar to our method. However, in some details, our method performs better than the DeepMicroGen.

\subsection{mbVDiT preserves the Pearson correlation coefficients between real data and imputed data for microbe-to-microbe}
To further validate the superior performance of our model. In this section, we calculated the Pearson correlation coefficients between each microbe in the imputed microbiome matrix for each method and the corresponding microbe in the actual expression matrix. After obtaining the correlation coefficient matrix, we sorted the PCC values for each microbe in descending order according to the mbVDiT method, and subsequent methods followed this order. 

\begin{figure}[htp]%
\centering
\includegraphics[width=0.5\textwidth]{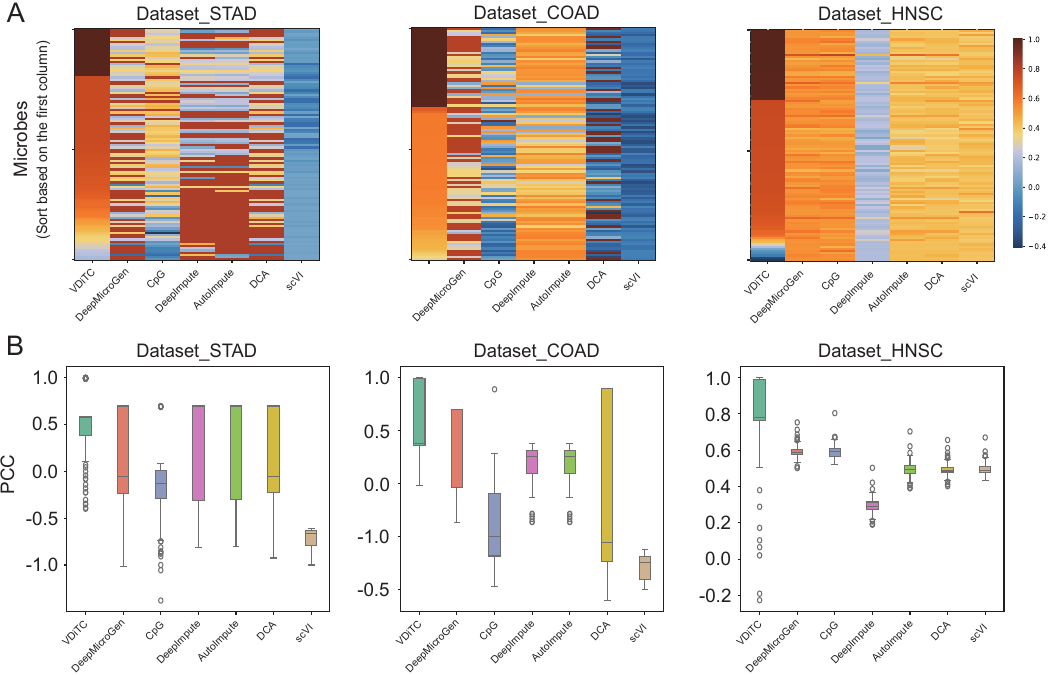}
\caption{(A) Heatmap of Pearson correlation coefficients between imputed and real microbiome data. (B) Box plot of Pearson correlation coefficients between imputed microbiome data and real microbiome data.}
\label{fig5}
\end{figure}

From the heatmap in section A of the figure (\autoref{fig5}), it can be seen that our method better preserves the original data distribution characteristics of each microbe after imputation compared to other methods. To facilitate a better visual comparison, we plotted a corresponding box plot using these PCC values (\autoref{fig5}). From section B of the figure, it can be further seen that our method generally outperforms the other methods.

\subsection{mbVDiT improves robustness in highly sparse microbiome data}
In microbiome data studies, the rate of missing data in patients with different cancer types may vary due to various reasons \cite{graham2013methods}. Data sparsity is defined as the percentage of zero elements in the expression matrix. To investigate the impact of different masking ratios on model performance, we applied three different masking ratios to the non-zero elements based on the original missing rate: 10\%, 30\%, and 50\%. Our method masks data in two stages: once on the original input data and again on the intermediate latent representation. In this, varying mask ratios are applied during the masking phase of the original input data, while the masking ratio in the second phase for the latent space is fixed at 30\%. Experimental results show that in our proposed method, different masking ratios have minimal impact on the results, further indicating that our method possesses strong stability (\autoref{tab6}).

\subsection{Ablation studies}
We conduct ablation experiment to verify the modules of mbVDiT in these aspects: (1) Different backbone networks; (2) Whether to include metadata as condition; (3) Whether to pre-training VAE.

\subsubsection{Different backbone networks}
In our model, the backbone network employs DiT Blocks, which are based on cross-attention mechanism. The purpose is to leverage the cross-attention mechanism to fuse different types of conditions, enabling the model to better generate new data. To validate the effectiveness of the cross-attention mechanism, we compared our method with the popular Unet network. 

The experimental results indicate that a backbone network predominantly based on attention mechanisms is more suitable for our specific task (\autoref{tab3}).

\begin{table}[htp]
\caption{Ablation experiments on different backbone networks. 
}
\label{tab3}
\begin{adjustbox}{width=0.49\textwidth}
\fontsize{10}{10}\selectfont    
\begin{tabular}{l|c|c|c}
\hline
\textbf{PCC $\uparrow$}             & \textbf{STAD}                               & \textbf{COAD}                               & \textbf{HNSC}                               \\ \hline
Backbone w/ Unet         & 0.523±0.037                                 & 0.597±0.066                                 & 0.538±0.011                                 \\
\textbf{Backbone w/ DiT} &  \textbf{0.634±0.032} &  \textbf{0.704±0.062} &  \textbf{0.626±0.060} \\ \hline
\hline
\textbf{Cosine $\uparrow$}          & \textbf{STAD}                               & \textbf{COAD}                               & \textbf{HNSC}                               \\ \hline
Backbone w/ Unet         & 0.755±0.021          &0.722±0.013          & 0.716±0.009          \\
\textbf{Backbone w/ DiT} &  \textbf{0.806±0.057} &  \textbf{0.791±0.080} &  \textbf{0.802±0.062} \\ \hline
\hline
\textbf{RMSE $\downarrow$}            & \textbf{STAD}                              & \textbf{COAD}                               & \textbf{HNSC}                               \\ \hline
Backbone w/ Unet         & 1.429±0.058                                & 1.238±0.038                                 & 1.246±0.050                                 \\
\textbf{Backbone w/ DiT} &  \textbf{1.32±0.053} &  \textbf{0.934±0.054} &  \textbf{1.155±0.027} \\ \hline
\hline
\textbf{MAE $\downarrow$}             & \textbf{STAD}                               & \textbf{COAD}                               & \textbf{HNSC}                               \\ \hline
Backbone w/ Unet         & 1.244±0.045                                 & 0.744±0.060                                 & 0.933±0.021                                 \\
\textbf{Backbone w/ DiT} & \textbf{0.956±0.050} & \textbf{0.530±0.070} &  \textbf{0.807±0.020} \\ \hline
\end{tabular}
\end{adjustbox}
\end{table}

\subsubsection{Whether to include metadata as condition}
In the field of bioinformatics, each dataset contains a wealth of information. In our method, we utilize various metadata from patients. To validate the effectiveness of this data, we conducted ablation experiments. 

Experimental data indicate that adding metadata to enrich the conditions of the diffusion model had slight improvement in four metrics. In conclusion, the experimental results also demonstrate that integrating various metadata to enrich conditions is effective, providing insights and assistance for subsequent experiments (\autoref{tab4}). This is a worthwhile area for further investigation. Our experiments directly validate that integrating various types of metadata is beneficial for deep learning models.

\begin{table}[htp]
\caption{Ablation experiments on whether the proposed mbVDiT with the condition (metadata). }
\label{tab4}
\begin{adjustbox}{width=0.49\textwidth}
\fontsize{8.5}{8.5}\selectfont         
\begin{tabular}{l|c|c|c}
\hline
\textbf{PCC $\uparrow$}         & \textbf{STAD}                               & \textbf{COAD}                               & \textbf{HNSC}                               \\ \hline
w/o Metadata         & 0.619±0.011                                 & 0.691±0.076                                 & 0.592±0.020                                 \\
\textbf{w/ Metadata} &  \textbf{0.634±0.032} & \textbf{0.704±0.062} & \textbf{0.626±0.060} \\ \hline
\hline
\textbf{Cosine $\uparrow$}      & \textbf{STAD}                               & \textbf{COAD}                               & \textbf{HNSC}                               \\ \hline
w/o Metadata         & 0.786±0.013                                 & 0.779±0.070                                 & 0.786±0.017                                 \\
\textbf{w/ Metadata} &  \textbf{0.806±0.057} & \textbf{0.791±0.080} &  \textbf{0.802±0.062} \\ \hline
\hline
\textbf{RMSE $\downarrow$}        & \textbf{STAD}                               & \textbf{COAD}                               & \textbf{HNSC}                               \\ \hline
w/o Metadata         & 1.331±0.060                                 & 0.963±0.054                                 & 1.173±0.032                                 \\
\textbf{w/ Metadata} & \textbf{1.320±0.053} & \textbf{0.934±0.054} &  \textbf{1.155±0.027} \\ \hline
\hline
\textbf{MAE $\downarrow$}        & \textbf{STAD}                               & \textbf{COAD}                               & \textbf{HNSC}                               \\ \hline
w/o Metadata         & 0.963±0.057                                 & 0.557±0.082                                 & 0.828±0.038                                 \\
\textbf{w/ Metadata} & \textbf{0.956±0.050} &  \textbf{0.530±0.070} &  \textbf{0.807±0.020} \\ \hline
\end{tabular}
\end{adjustbox}
\end{table}

\subsubsection{Whether to pre-train VAE}
Due to the limited amount of data in individual cancer datasets, training on a single dataset cannot achieve the expected results for the model. Therefore, we use pre-training method to improve model performance. By leveraging datasets from other types of cancers, we learn a shared weight matrix and transfer this matrix to the VAE module in mbVDiT as initial weights. In order to verify that pre-training can improve our model performance, we conducted comparative experiments on the VAE module of mbVDiT.  

The results (\autoref{tab5}) showed that using pre-training significantly improved the performance of the model. This also proves that pre-training can effectively address the objective factor of a small sample size in the datasets.

\begin{table}[htp]
\caption{Ablation experiments on whether pre-training VAE module. 
}
\label{tab5}
\begin{adjustbox}{width=0.49\textwidth}
\fontsize{8}{8}\selectfont    
\begin{tabular}{l|c|c|c}
\hline
\textbf{PCC $\uparrow$}           & \textbf{STAD}                               & \textbf{COAD}                               & \textbf{HNSC}                               \\ \hline
w/o Pre\_train         & 0.518±0.011                                 & 0.613±0.065                                 & 0.589±0.014                                 \\
\textbf{w/ Pre\_train} &  \textbf{0.634±0.032} &  \textbf{0.704±0.062} &  \textbf{0.626±0.060} \\ \hline
\hline
\textbf{Cosine $\uparrow$}        & \textbf{STAD}                               & \textbf{COAD}                               & \textbf{HNSC}                               \\ \hline
w/o Pre\_train         & 0.648±0.009                                 & 0.737±0.040                                 & 0.741±0.013                                 \\
\textbf{w/ Pre\_train} &  \textbf{0.806±0.057} &  \textbf{0.791±0.080} &  \textbf{0.802±0.062} \\ \hline
\hline
\textbf{RMSE $\downarrow$}          & \textbf{STAD}                              & \textbf{COAD}                               & \textbf{HNSC}                               \\ \hline
w/o Pre\_train         & 1.442±0.013                                & 0.996±0.015                                 & 1.276±0.026                                 \\
\textbf{w/ Pre\_train} &  \textbf{1.32±0.053} &  \textbf{0.934±0.054} &  \textbf{1.155±0.027} \\ \hline
\hline
\textbf{MAE $\downarrow$}           & \textbf{STAD}                               & \textbf{COAD}                               & \textbf{HNSC}                               \\ \hline
w/o Pre\_train         & 1.045±0.014                                 & 0.561±0.011                                 & 0.870±0.010                                 \\
\textbf{w/ Pre\_train} &  \textbf{0.956±0.050} &  \textbf{0.530±0.070} &  \textbf{0.807±0.020} \\ \hline
\end{tabular}
\end{adjustbox}
\end{table}

\begin{table}[htp]
\caption{Robustness experiments of model performance under different masked ratios.}
\label{tab6}
\begin{adjustbox}{width=0.49\textwidth}
\fontsize{14}{15}\selectfont    
\begin{tabular}{l|cccccc}
\hline
                                & \multicolumn{3}{c}{\textbf{PCC $\uparrow$}}                                   & \multicolumn{3}{c}{\textbf{Cosine $\uparrow$}}                                \\ \cline{2-7} 
\multirow{-2}{*}{\textbf{Rate}} & \textbf{10\%} & \textbf{30\%}                      & \textbf{50\%} & \textbf{10\%} & \textbf{30\%}                      & \textbf{50\%} \\ \hline
STAD                            & 0.619±0.010   &  0.634±0.032 & 0.609±0.016   & 0.794±0.006   &  0.806±0.057 & 0.787±0.019   \\
COAD                            & 0.693±0.010   &  0.704±0.062 & 0.703±0.065   & 0.791±0.007   &  0.791±0.080 & 0.796±0.049   \\
HNSC                            & 0.628±0.019   &  0.626±0.060 & 0.613±0.007   & 0.793±0.026   & 0.802±0.062 & 0.796±0.036   \\ 
\hline
\hline
                                & \multicolumn{3}{c}{\textbf{RMSE $\downarrow$}}                                  & \multicolumn{3}{c}{\textbf{MAE $\downarrow$}}                                   \\ \cline{2-7} 
\multirow{-2}{*}{\textbf{Rate}} & \textbf{10\%} & \textbf{30\%}                      & \textbf{50\%} & \textbf{10\%} & \textbf{30\%}                      & \textbf{50\%} \\ \hline
STAD                            & 1.273±0.029   &  1.320±0.053  & 1.306±0.065   & 0.910±0.019   &  0.956±0.050 & 0.935±0.067   \\
COAD                            & 0.868±0.050   &  0.934±0.054 & 0.937±0.049   & 0.507±0.008   &  0.530±0.070 & 0.526±0.039   \\
HNSC                            & 1.159±0.028   &  1.155±0.027 & 1.161±0.024   & 0.812±0.021   & 0.807±0.020 & 0.816±0.003   \\ \hline
\end{tabular}
\end{adjustbox}
\end{table}

\section{CONCLUSIONS}
In this paper, we proposed a masked conditional diffusion model with variational autoencoder (mbVDiT) for microbiome data denoising and imputation. By masking part of the latent representation and utilizing the remaining portion and patient metadata as condition, the diffusion model becomes well-guided and controllable throughout the reverse process. In addition, mbVDiT uses a VAE-based pre-trained model to integrate other public microbiome datasets to enhance its performance. Experiments on three TCGA microbiome data showed that our approach achieves the best overall performance compared to other baseline methods.

\section*{Acknowledgment}
The work was supported in part by the National Natural Science Foundation of China (62262069), in part by the Yunnan Fundamental Research Projects under Grants (202201AT070469) and the Yunnan Talent Development Program - Youth Talent Project.

\small
\bibliography{Ref.bib} 
\bibliographystyle{IEEEtran}  
\end{document}